\documentstyle[epsfig]{mn2e}
\begin{document}
\input BoxedEPS.tex
\newcommand{\aip}{{\small ${\cal AIPS}$}}
\newcommand{\gtsim}{\mbox{{\raisebox{-0.4ex}{$\stackrel{>}{{\scriptstyle\sim}}
$}}}}
\newcommand{\ltsim}{\mbox{{\raisebox{-0.4ex}{$\stackrel{<}{{\scriptstyle\sim}}
$}}}}
\newcommand{\s}{$\stackrel{\rm s}{.}$}
\newcommand{\h}{$^{\rm h}$}
\newcommand{\m}{$^{\rm m}$}
\newcommand{\pp}{$\stackrel{\prime\prime}{.}$}
\newcommand{\de}{$^{\circ}$}
\newcommand{\p}{$^{\prime}$}
\newcommand{\arc}{$^{\prime\prime}$}
\newcommand{\marc}{^{\prime\prime}}
\newcommand{\rs}{{\em $r_s$}}
\newcommand{\DPM}{{\em DPM}}
\newcommand{\alf}{{\displaystyle\biggl({\nu_{\rm h} \over \nu_{\rm l}}\biggr)^{\alpha}} }

\newcommand{\figstart}[1]
    { \begin{figure}[htb]
      \begin{picture}(0,#1) }
\newcommand{\figend}[4]
    { \end{picture}
      \special{#1}
      \caption[#2]{#3}
      \label{#4}
      \end{figure} }
\newcommand{\fig}[5]
    { \figstart{#1}
      \figend{#2}{#3}{#4}{#5} }
\newcommand{\bHS}{\beta_{\mbox{\scriptsize HS}}}
\newcommand{\bBF}{\beta_{\mbox{\scriptsize BF}}}
\newcommand{\nT}{\nu_{\mbox{\scriptsize T}}}
\newcommand{\et}{E_{\mbox{\scriptsize T}}}
\newcommand{\nTn}{\nu_{\mbox{\scriptsize Tn}}}
\newcommand{\nTf}{\nu_{\mbox{\scriptsize Tf}}}
\newcommand{\tn}{\tau_{x\mbox{\scriptsize n}}}
\newcommand{\tf}{\tau_{x\mbox{\scriptsize f}}}
\newcommand{\xn}{x_{\mbox{\scriptsize n}}}
\newcommand{\xf}{x_{\mbox{\scriptsize f}}}
\newcommand{\yn}{y_{\mbox{\scriptsize n}}}
\newcommand{\yf}{y_{\mbox{\scriptsize f}}}
\newcommand{\lln}{l_{\mbox{\scriptsize n}}}
\newcommand{\llf}{l_{\mbox{\scriptsize f}}}
\newcommand{\Dn}{f(\Delta_{\mbox{\scriptsize n}})}
\newcommand{\Df}{f(\Delta_{\mbox{\scriptsize f}})}
\newcommand{\B}{\mbox{$B$}}
\newcommand{\Bo}{\mbox{$B$}_{0}}

\SetRokickiEPSFSpecial
\HideDisplacementBoxes

%%%%%%%%%%%%%%% MODIFICATION HISTORY %%%%%%%%%%%%%%%%%%%%
%%%%%
%%%%%  
%%%%%%%%%%%%%%%%%%%%%%%%%%%%%%%%%%%%%%%%%%%%%%%%%%%%%%%%%

\title[A model for the infrared emission of FSC 10214+4724]
{A model for the infrared emission of FSC 10214+4724}
  \author[Efstathiou]
  {Andreas Efstathiou\\
School of Computer Science \& Engineering, Cyprus College, 
Diogenes Street, Engomi, 1516 Nicosia, Cyprus.\\ 
}
\maketitle
\begin{abstract}

 A model for the infrared emission of the high redshift ultraluminous
infrared galaxy FSC 10214+4724 is presented. The model assumes three components
of emission: a dusty torus viewed edge-on, clouds that are associated with the
narrow-line region and a highly obscured starburst. It is demonstrated that the
presence of clouds in the narrow-line region, with a covering factor of 17\%,
 can explain why the mid-infrared spectrum of FSC 10214+4724 shows
a silicate feature in emission despite the fact that its torus is viewed edge-on. 
It is also shown that the same model, but with the torus viewed face-on, predicts
a spectrum with silicate emission features that is characteristic
of the spectra of quasars recently observed with $Spitzer$.

\end{abstract}
\begin{keywords}
galaxies:$\>$ active -
galaxies:$\>$ individual (FSC 10214+4724) -
infrared:$\>$galaxies -
dust:$\>$ -
radiative transfer:$\>$
\end{keywords}

%\large

\section{Introduction}
\label{}

The z=2.285 galaxy FSC10214+4724 was discovered 15 years ago by Rowan-Robinson
 et al. (1991). At the time of its discovery it was thought to be one of the most
luminous objects in the Universe. It was later found to be gravitationally
lensed by a factor 10-100 (Broadhurst \& Lehar 1995, Graham \& Liu 1995,
Serjeant et al. 1995, Eisenhardt et al. 1996). It is still the best
studied ultraluminous infrared galaxy at that redshift. Rest frame optical-UV
spectroscopy and spectropolarimetry showed that FSC 10214+4724 exhibits
similar characteristics to Seyfert 2 galaxies (Elston et al. 1994,
Lawrence et al. 1993, Goodrich et al. 1996). A large quantity of molecular
gas is suggested by CO observations (Brown \& Vanden Bout 1991, Solomon
et al. 1992, Scoville et al. 1995). The object is also bright in the
submillimetre. It is therefore likely that FSC 10214+4724 is similar to 
some local ultraluminous infrared galaxies which harbour simultaneously 
a luminous starburst and an AGN (e.g. Genzel et al. 1998).

Radiative transfer models of emission from dust in a starburst or an AGN
were used in an attempt to understand the origin of the infrared emission of
FSC10214+4724. Rowan-Robinson et al. (1993) showed that good fits to the 
overall spectral energy distribution (SED) could be obtained with both a starburst and
an AGN model. The dust density in the latter falls off as $r^{-1}$ and
varies exponentially with azimuthal angle (Efstathiou \& Rowan-Robinson 1995).
Green \& Rowan-Robinson (1996) concluded that starburst or AGN activity alone
can not explain the SED of FSC 10214+4724 but instead a combination of
the two is needed. Granato, Danese \& Franceschini (1996) found that a
good fit to the SED of FSC 10214+4724 and other hyperluminous infrared
galaxies could be obtained with pure dust-enshrouded AGN models.
The limited wavelength coverage in the infrared is the main reason why
modeling attempts in the mid-90s remained inconclusive.

Recently Teplitz et al. (2006) presented a mid-infrared spectrum for
FSC 10214+4724 which was obtained with the Infrared Spectrograph onboard 
$Spitzer$ (Houck et al. 2004). This spectrum places strong constraints on
models of the infrared emission of this object. What is interesting about the
spectrum presented by Teplitz et al. is that the silicate feature appears in 
emission. This is surprising given the type 2 nature of the AGN in FSC 10214+4724.
Local examples of type 2 AGN (e.g. NGC1068) show silicate features in absorption
in agreement with radiative transfer models of dusty tori viewed almost edge-on
(Pier \& Krolik 1992, Granato \& Danese 1994, Efstathiou \& Rowan-Robinson 1995).
The spetrum presented by Teplitz et al. shows no signs of PAH features except
possibly a weak 6.2$\mu m$ feature. This implies that the starburst if present
is weak or highly obscured.

In this letter radiative transfer models of dusty AGN and dusty
starbursts are used to fit the now detailed SED of FSC 10214+4724. Section 2
describes briefly the radiative transfer models to be used. In section 3 
the results are presented and briefly discussed. A flat Universe is assumed
with $\Lambda =0.73$ and H$_0$=71Km/s/Mpc.

\section{Radiative transfer models}

\subsection{Starburst models}

Radiative transfer models for starbursts were presented by Rowan-Robinson
\& Crawford (1989), Rowan-Robinson \& Efstathiou (1993), Kr\"ugel \& Siebenmorgen
(1994), Silva et al. (1998), Efstathiou et al. (2000), Siebenmorgen et al (2001),
Tagaki et al. (2003), Dopita et al. (2005). For this study  the models of Efstathiou
 et al. (2000) which were extended to higher optical depths by Efstathiou \&
Siebenmorgen (2005) will be used. The models incorporate the stellar population synthesis
models of Bruzual \& Charlot (1993), a simple model for the evolution of the
molecular clouds that constitute the starburst and detailed radiative transfer
that includes the effect of transiently heated particles/PAHs (Siebenmorgen \&
Kr\"ugel 1992). In Fig. 1 the model predictions for starbursts of different optical
 depths are compared. As expected the ratio of far- to mid-infrared luminosity
increases with optical depth whereas the submillimeter spectrum remains largely
unaffected. The model with $\tau_V =50$ predicts an SED similar to that of the nearby
starburst galaxy M82 whereas the most optically thick model gives an SED similar
to that of the heavily obscured ultraluminous infrared galaxy Arp220.

\begin{figure}
\epsfig{file=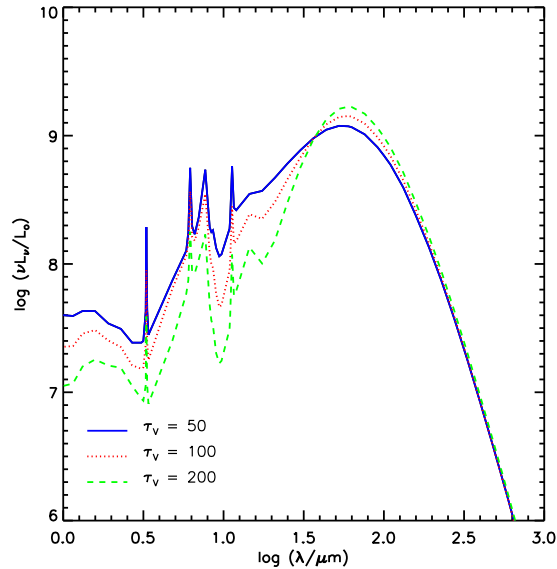, angle=0, width=8cm}
\caption{
Spectral energy distributions of  Efstathiou et al. starbursts
in which the star formation rate decays exponentially with
time (time constant 30Myrs). The SED is shown for three different
values of the initial optical depth in the V band of the molecular
clouds that constitute the starburst. The initial star formation rate 
is assumed to be 1 $M_\odot/yr$ and the starburst age is assumed to be 40Myrs.
}
\end{figure}

\subsection{AGN torus models}

Models of the infrared emission from dusty tori and discs were presented
by Pier \& Krolik (1992, 1993), Granato \& Danese (1994), Efstathiou \&
Rowan-Robinson (1995), van Bemmel \& Dullemond (2003), Dullemond \& van
Bemmel (2005). Discussion centred on the problem of minimizing the silicate
emission features at 10$\mu m$ which were not observed in the ground-based 
8-13$\mu m$ observations of predominantly Seyfert galaxies by Roche et al. (1991).
The non-detection of silicates was attributed either to the torus geometry
(Pier \& Krolik 1992, Efstathiou \& Rowan-Robinson 1995) or to the destruction
of silicate dust in the inner torus (Granato \& Danese 1994). Rowan-Robinson (1995)
and more recently Nenkova et al. (2002) proposed that the non-detection of silicates
is due to the clumpiness of the dust but Dullemond \& van Bemmel (2005) showed 
that clumpiness does not necessarily eliminate the silicate features.

Recent results from $Spitzer$ (Siebenmorgen et al. 2005, Hao et al. 2005) 
showed that quasars emit emission features both at 10 and 18$\mu m$. Although
a detection of silicate emission in a LINER was also reported by Sturm et al. (2005)
it seems that silicate emission is found mainly in high luminosity AGN. This is
an observational fact that models of the infrared emission of AGN must therefore
explain. The models should also explain the fact that some luminous type 2 AGN 
like FSC 10214+4724 (see also Sturm et al 2006) also show emission features at
 10$\mu m$.

Another observational fact  is that about half of the mid-infrared
emission of the prototypical Seyfert 2 galaxy NGC1068 is extended and coincides
with the narrow line region (e.g Cameron et al 1993, Braatz et al 1993, Galliano et al.
2005). Efstathiou, Hough \& Young (1995) presented a model for the nuclear spectrum of
NGC1068 that involved a tapered disc (Efstathiou \& Rowan-Robinson 1995) and optically
thin dust in a conical region. The latter may actually reside in the narrow line region 
(NLR) clouds. The model considered in this letter for the AGN contribution to the spectrum of 
FSC 10214+4724 is a combination of the tapered disc models of Efstathiou \& 
Rowan-Robinson (1995) and a simple model for the emission of clouds that are
associated with the NLR. A tapered disc has a thickness that increases linearly with
distance from the central source in the inner part of the disc but assumes a constant
height in the outer disc. Although the models of Efstathiou \& Rowan-Robinson do not
consider clumping of the dust in clouds they can be considered as a reasonable
approximation to the dust distribution in AGN if the distance between clouds is much
less than the size of the torus.

\section{Results and discussion}

 Detailed modeling of the emission of NLR clouds
that are illuminated by a central source is beyond the scope
of this letter. The emission of such clouds is approximated  with that
of spherical clouds of dust at a constant temperature. It is assumed that
the density in the cloud is uniform and the optical depth in the V band
from the center to the surface is 10. The same grain mixture used for the
dusty starburst radiative transfer calculations is assumed.
A good fit to the $Spitzer$ spectrum can be obtained by assuming 
that the NLR dust is concentrated in clouds at a temperature of
200 and 610K with the 200K cloud(s) being about an order of magnitude more luminous.
This is broadly in agreement with the findings of Teplitz et al. (2006). The
starburst model assumed is the $\tau_V = 200$ model shown in Fig. 1.
 
The model assumed for the torus is the most geometrically thin of the models
considered by Efstathiou \& Rowan-Robinson ($\Theta_1=30^o$). This implies
a torus half opening angle of 60$^o$ which, however, is very poorly constrained.
The model also assumes a ratio of inner to outer disc radius of 0.05, a ratio of height
to outer radius of 0.5 and an equatorial optical depth at 1000\AA~ of 1000. 
It is assumed that the tapered dusty disc in FSC 10214+4724 is viewed edge-on. 
This means that its mid-infrared radiation is heavily suppressed. The sum of the
emission of the NLR dust and the disc is shown in Fig. 2 and 3 whereas the
individual spectra are shown in Fig. 4. Note that because the starburst
is very optically thick it does not make a significant contribution to
the mid-infrared flux. This can explain why there is no evidence of PAH
features in the $Spitzer$ spectrum published by Teplitz et al. except for
a weak 6.2$\mu m$ feature that is reproduced by the model.

 As can be seen from the plot of the edge-on and face-on spectra emitted
by the torus in Fig. 4, the AGN luminosity is highly anisotropic. To obtain
the total luminosity one needs to multiply the `apparent' torus luminosity 
with the anisotropy correction factor $A$ which is defined to be

 $$  A(i) = {{ \int_0^{\pi/2} S(i')\ \sin i' \ di'} \over
      {S(i)}}  $$
where $S(i)$ is the bolometric emission at angle $i$. For the particular
tapered disc model and inclination, $A$ is found to be 18.1. 
 The emission by the NLR dust is assumed to be isotropic. With this
correction the AGN luminosity is predicted to be $6.9 \times 10^{14} L_\odot$. The
luminosity of the 200K NLR dust is predicted to be $1.0 \times 10^{14} L_\odot$
or 15\% of the total AGN luminosity. As mentioned above the luminosity of
the 610K dust is about an order of magnitude lower or 2\% of the total. 
As the NLR clouds are assumed to be optically thick, the fraction of total AGN
luminosity emitted by the 200 and 610K dust is also the covering factor of the NLR
 dust. The starburst luminosity is predicted to be $7.6 \times 10^{13} L_\odot$.
Note that all of the quoted luminosities need to be reduced by a factor  10-100 due to
gravitational lensing. Lens models suggest that the optical emission may be
more highly magnified than the infrared emission that dominates the bolometric
luminosity. Eisenhardt et al. (1996) estimate a magnification factor for the 
infrared emission of 30 which implies that the AGN luminosity is of the order 
of $2.3 \times 10^{13} L_\odot$. Note, however, that since the emission by the
NLR dust and the torus does not arise from the same region it is possible that their
magnification will be different.

\begin{figure}
\epsfig{file=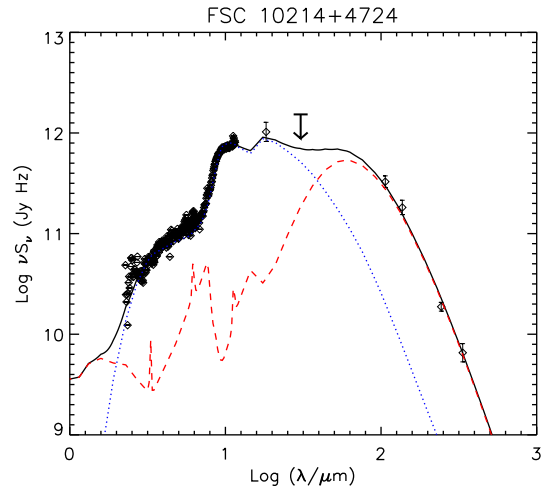, angle=0, width=8cm}
\caption{
Fit to the spectral energy distribution of FSC 10214+4724. The dotted and
dashed lines show the contributions of the AGN and starburst respectively
to the total emission (solid line). Data from Teplitz et al (2006), IRAS,
Rowan-Robinson et al. (1993).
}
\end{figure}

\begin{figure}
\epsfig{file=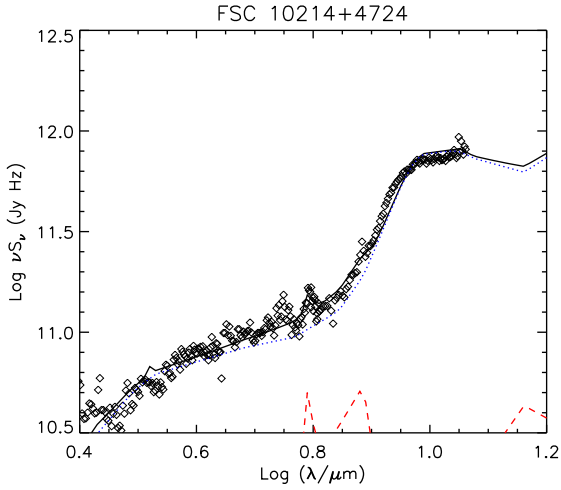, angle=0, width=8cm}
\caption{
As Fig. 2 but showing the fit to the mid-infrared spectrum in more detail.
}
\end{figure}

\begin{figure}
\epsfig{file=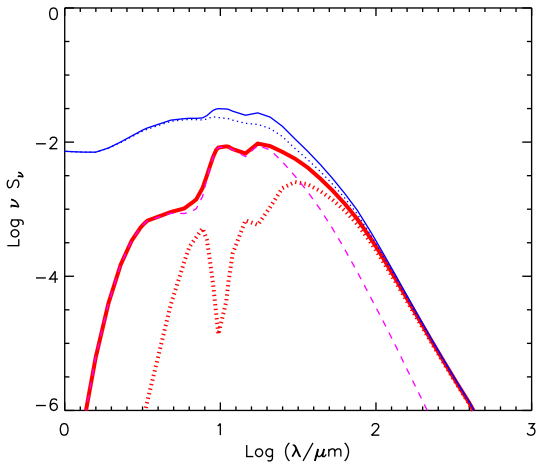, angle=0, width=8cm}
\caption{
Comparison of the edge-on and face-on views of FSC 10214+4724
according to the model presented here. The thick (red) dotted line gives the spectrum
emitted by the edge-on torus and the dashed line (magenta) gives the emission
of the dust in the cones. The thick (red) solid line gives the total emission in
the edge-on case. The thin (blue) dotted line gives the predicted spectrum
for a face-on view of the torus and the thin (blue) solid line gives
the sum of that emission to the emission of the dust in the cones which is
assumed to radiate isotropically. The vertical scale is arbitrary.
}
\end{figure}

As it is demonstrated in Figs. 2, 3 and 4  the presence of dust in the 
narrow-line region of an AGN nucleus, with a covering factor of 17\%, 
can explain the fact that silicate features appear in emission both in 
luminous type 2 and type 1 AGN. The presence of these features in type 
2 AGN is entirely consistent with the unified scheme (Antonucci 1993)
as a dust torus that makes an insignificant contribution in the mid-infrared
 may still be present. The anisotropy of the emission from such a torus is so high that
when the same system is viewed face-on the torus emission dominates
and the effect of the presence of the NLR dust is to make the
silicate emission features more prominent. The face-on spectrum shown
in Fig. 4 is similar to the spectra of quasars recently observed
with $Spitzer$.

A question that naturally arises is why lower luminosity type 2 AGN do not
show the same effect. There are probably two reasons for this: firstly in lower 
luminosity type 2s the ratio of the luminosity of the acompanying starburst to
 that of the AGN is higher so what we see in the mid-infrared is the complex of PAH
features and silicate absorption features emitted by the starburst.
Secondly the torus opening angle in lower luminosity AGN may be smaller
and therefore the fraction of the source luminosity that is intercepted
by NLR dust is smaller. This was already suggested by Lawrence (1991)
who estimated the statistics of type 1 and type 2 AGN from different surveys. As
proposed by Lawrence this dependence on luminosity may arise if
the thickness of the torus is independent of luminosity and its
inner radius is controlled by dust sublimation.

\section*{Acknowledgments}

 I thank Harry Teplitz for providing the spectrum of FSC 10214+4724 in
electronic form and Ralf Siebenmorgen for useful discussions. I also
thank the referee Andy Lawrence for his comments and suggestions that
led to a significant improvement of this letter.

\end{document}